\documentclass[letterpaper,12pt,hidelinks]{article}
\usepackage{amsfonts, amsmath, amsthm, graphicx, float,
  subcaption, enumerate, color, natbib, algorithm,
  algorithmic, url, hyperref, amssymb, bm, mathtools}
\usepackage[shortlabels]{enumitem}
\usepackage[margin=1in]{geometry}
\usepackage[onehalfspacing]{setspace}
\DeclarePairedDelimiter\ceil{\lceil}{\rceil}

\graphicspath{{figures/}}

\newcommand{\M}{\bm{M}}
\renewcommand{\S}{\bm{S}}

\renewcommand{\u}{\bm{u}}

\newcommand{\x}{\bm{x}}

\newcommand{\y}{\bm{y}}
\newcommand{\z}{\mathbf{z}}
\newcommand{\bdelta}{\bm{\delta}}
\newcommand{\cz}{\mathcal{Z}}
\newcommand{\cx}{\mathcal{X}}

\newcommand{\bbeta}{\boldsymbol{\beta}}
\newcommand{\hbeta}{\hat{\boldsymbol{\beta}}}
\newcommand{\bSigma}{\boldsymbol{\Sigma}}

\newcommand{\op}{o_{P}(1)}
\newcommand{\Op}{O_{P}(1)}

\newcommand{\Fn}{\mathcal{D}_N}

\newcommand{\tp}{^{\rm T}}

\newcommand{\Exp}{\mathbb{E}}
\newcommand{\Var}{\mathbb{V}}

\newtheorem{theorem}{Theorem}
\theoremstyle{remark}
\newtheorem{remark}{{\bf Remark}}

\newcommand{\sumrn}{\sum_{i=n-r+1}^{n}}
\newcommand{\sumrbN}{\sum_{i=N-r_B+1}^{N}}
\newcommand{\sumb}{\sum_{b=1}^{B}}
\newcommand{\vr}{\mathrm{var}}
\newcommand{\sump}{\sum_{j=1}^{p}}
\newcommand{\ds}{\mathcal{D}_S}
\newcommand{\dsB}{\mathcal{D}_{{\tiny{\mathrm{BS}}}}}
\newcommand{\dbs}{\mathcal{D}_{S}^{(b)}}
\newcommand{\sumk}{\sum_{i=1}^{k}}

\newcommand{\sumN}{\sum_{i=1}^{N}}

\newcommand{\sumr}{\sum_{i=1}^{r}}%
\newcommand{\sumkb}{\sum_{i=1}^{k_B}}
\newcommand{\sumrb}{\sum_{i=1}^{r_B}}%

\begin{document}
\title{Divide-and-Conquer Information-Based Optimal Subdata Selection
  Algorithm
  \footnote{This work was supported by NSF grant 1812013, a
    UCONN REP grant, and a GPU grant from NVIDIA Corporation.}}
\author{ HaiYing Wang\\
  Department of Statistics, University of Connecticut\\
  haiying.wang@uconn.edu}
\date{}

\maketitle
\begin{abstract}
  The information-based optimal subdata selection (IBOSS) is a computationally efficient method to select informative data points from large data sets through processing full data by columns. However, when the volume of a data set is too large to be processed in the available memory of a machine, it is infeasible to implement the IBOSS procedure. This paper develops a divide-and-conquer IBOSS approach to solving this problem, in which the full data set is divided into smaller partitions to be loaded into the memory and then subsets of data are selected from each partitions using the IBOSS algorithm. We derive both finite sample properties and asymptotic properties of the resulting estimator. Asymptotic results show that if the full data set is partitioned randomly and the number of partitions is not very large, then the resultant estimator has the same estimation efficiency as the original IBOSS estimator. We also carry out numerical experiments to evaluate the empirical performance of the proposed method.
\end{abstract}

\section{Introduction}

In big data analysis, a common challenge is that available computing facilities are inadequate to meet the computational needs. To overcome this challenge, there are two fundamental approaches: one is the divide-and-conquer approach in which a large data set is divided into smaller partitions and results are calculated from each partition and then combined; the other approach is to use a subset of the full data as a surrogate, and use calculation results from the subdata to approximate the full data calculation results. 

The divide-and-conquer approach naturally fits into parallel and distributed computing systems, and it is effective to deal with the challenge that the data volume is too large to be loaded into the memory although it may not always reduce the computing time with a single processor. Most distributed computational platforms are expensive to access and they are not suitable for daily routine usages such as exploratory data analysis and pilot model prototyping.
The divide-and-conquer method has attracted many researchers in machine learning and statistics, leading to key advancements in \cite{LinXie2011,
  jordan2012divide, chen2014split, shang2017computational, battey2018distributed}, among others. 
In the streaming setting where data blocks are accessible sequentially for only one time, the online updating method has been developed \citep{schifano2016online, xue2018online}. In order to produce an estimator that preserves the same convergence rate as the full data estimator, a key requirement for the divide-and-conquer method is that the number of partitions cannot be too large \citep{shang2017computational}.

The approach of using a subset of the full data is an effective method to reduce the computational burden, and it is often the only way to extract useful information from massive data sets when available computing power is limited. An advantage of this approach is that once a subset of data is taken, thorough analysis can often be performed on a regular computer such as a laptop. This is important in statistics because data analysis is not just computing. 
Note that the primary goal of selecting an informative subset 
from the full data agrees with the basic motivation of optimal design of experiments in which one wants to get as much information as possible with a fixed design budget. The major challenge of subdata selection from big data is how to identify informative data points computationally fast in order to maintain the computational advantage in the whole procedure of data analysis. 
For this purpose, statistical leverage scores are often used to define subsampling probabilities in linear regression \citep{Drineas:12,PingMa2015-JMLR}. In the context of logistic regression, \cite{WangZhuMa2017} derived the optimal subsampling probabilities that minimize the asymptotic variance of the subsampling estimator under the A- and L-optimality criteria, 
and \cite{wang2018more} further developed a more efficient estimation approach based on optimal subsamples. \cite{yao2018optimal} extended this technique to include the softmax regression.

The aforementioned work uses random subsampling to take subsets of full data, and \cite{WangYangStufken2018} showed that an estimator obtained from this approach for linear regression has a variance that is bounded from below by a term that is proportional to the inverse of the subset sample size, i.e., the estimator will not converge to the true parameter if the subset sample size is fixed no matter how fast the full data sample size goes to infinity. From the characterization of the optimal subdata under the D-optimality criterion, they further develop a computationally efficient IBOSS algorithm. The resulting estimator from this algorithm converges to the true parameter even if the subdata sample size is fixed as long as the full data sample size gets large and the supports of the covariate distributions are unbounded.
The IBOSS algorithm selects data points by examining each column of the design matrix, so it is infeasible to apply if the data volume is too large to be loaded into the random-access memory (RAM). This paper combines the divide-and-conquer method and the IBOSS algorithm to solve this issue. The basic approach is to divide a large data set into smaller partitions so that each partition can be loaded into the RAM and the IBOSS algorithm is applied on the smaller partitions. In practice, big data are seldom stored in a single file; they are often stored in multiple data blocks. The proposed method fits this scenario naturally. 
We will investigate both finite sample properties and asymptotic properties of this procedure, and use numerical experiments to evaluate its empirical performance.

The rest of the paper is organized as the following. We present notations for the model setup and introduce the IBOSS method in Section~\ref{sec:framework}. The main theoretical results on the divide-and-conquer IBOSS algorithm will be presented in Section~\ref{sec:divide-conquer-iboss}. Section~\ref{sec:numer-exper} evaluates the empirical performance of the divide-and-conquer IBOSS algorithm using numerical examples. Section~\ref{sec:concluding-remarks} concludes the paper and all proofs are provided in the Appendix. 

\section{IBOSS framework and detailed algorithm}
\label{sec:framework}
Assume that the full data $\Fn=\{(\z_1, y_1), ..., (\z_N, y_N)\}$ satisfy the following linear regression model
\begin{equation}\label{eq:14}
  y_i=\beta_0+\sump z_{ij}\beta_j+\varepsilon_i,\ \quad i=1,...,N,
\end{equation}
where $y_i$ are the responses, $\z_i=(z_{i1}$, ..., $z_{ip})\tp$ are the
covariates, $\beta_0$ and $\beta_1, ..., \beta_p$ are the intercept
and slope parameters, respectively, and $\varepsilon_i$ are model errors. For easy
of presentation, we define some notations. Let $\x_i=(1, \z_i)$  
and 
$\cz
=(\z_1\tp, ..., \z_N\tp)\tp$. 
Denote
$\bbeta_1=(\beta_1,\beta_2, ..., \beta_p)\tp$ and
$\bbeta=(\beta_0, \bbeta_1\tp)\tp$. 
  Here we assume that the
error terms $\varepsilon_i$ are uncorrelated and satisfy
$\Exp(\varepsilon_i)=0$ and $\Var(\varepsilon_i)=\sigma^2$. Without loss of generality, we assume that $\sigma^2=1$ when discussing the selection of subdata sets. 

To estimate the unknown $\bbeta$, the best linear unbiased estimator (BLUE) is the least squares estimator. With the full data
$\Fn$, it has an expression of
\begin{equation*}
  \hbeta_{\mathrm{full}}=\left(\sumN \x_i\x_i\tp\right)^{-1}\sumN \x_iy_i,
\end{equation*}
with variance-covariance matrix
\begin{equation*}
  \Var(\hbeta_{\mathrm{full}}|\cz)=\M_{\mathrm{full}}^{-1}
  =
  \bigg(\sumN \x_i\x_i\tp\bigg)^{-1},
\end{equation*}
where $\M_{\mathrm{full}}$ is the observed Fisher information matrix of $\bbeta$ for the full data if $\varepsilon_i$ are normally distributed. 

When the sample size $N$ of the full data set is too large, full data analysis may take too long to afford. To extract some useful information from the data in time with limited computing resources, a subset of the full data can be selected and thoroughly analyzed. For this purpose, \cite{WangYangStufken2018} proposed the IBOSS method. We summarize the motivation and procedure in the following. 

Use $\bdelta=\{\delta_1, \delta_2, ..., \delta_N\}$ to indicate a subset of data where $\delta_i=1$
if $(\z_i, y_i)$ is included in the subset and $\delta_i=0$
otherwise. Here $\bdelta$ may depend on $\cz$ but it does not depend on $\y$, so the subdata selection rule is ancillary to the regression parameter, and the least squares estimator is still the BLUE based on the subset of the full data. With this notation, the information matrix for $\bbeta$ based on a subset of the full data indexed by $\bdelta$ is
\begin{equation*}
  \M_{\bdelta}=
  \sumN\delta_i\x_i\x_i\tp. 
\end{equation*}
To extract the maximum amount of information from the full data with a
fixed subset sample size $k$ so that $|\bdelta|=\sumN\delta_i=k$,
\cite{WangYangStufken2018} proposed to select the subset of data that maximizes $\M_{\bdelta}$ with respect to $\bdelta$ under the the D-optimality criterion in optimal design of experiments, namely, to find
\begin{equation*}
  \bdelta^{opt}_{\mathrm{D}}=
  \underset{\bdelta:|\bdelta|=k}{\arg\max}
  \left|\sumN\delta_i\x_i\x_i\tp\right|.
\end{equation*}
In general, there is no analytical solution to $\bdelta^{opt}_{\mathrm{D}}$, and
numerical search for an exact solution is computationally NP-hard. \cite{WangYangStufken2018} derived an upper bound for
$|\M_{\bdelta}|$, and then proposed an algorithm to approximate it. To be specific, they showed that for any $\bdelta$ with $|\bdelta|=k$,
\begin{align}\label{eq:15}
  |\M_{\bdelta}|\le\zeta_N
  :=\frac{k^{p+1}}{4^p}\prod_{j=1}^p(z_{(N)j}-z_{(1)j})^2,
\end{align}
where $z_{(N)j}$ and $z_{(1)j}$ are the sample maximum and sample minimum, respectively, for the $j$-th covariate. 
Although the upper bound $\zeta_N$ is typically unachievable, it indicates that the more informative data points are in the tail regions of the covariates. This motivated the IBOSS algorithm to approximate the upper bound, and here is a summary of the IBOSS procedure. Assume that $r=k/(2p)$ is an integer. For each covariate, 
using a partition-based selection algorithm
\citep{Martinez2004}, select $r$ data points with the smallest
$z_{ij}$ values and $r$ data points with the largest $z_{ij}$ values
among the data points that are not yet included in the subsample. The data points obtained according to the $p$ covariates are then combined to form a subdata of size $k$ for statistical analysis. 

The original IBOSS paper by \cite{WangYangStufken2018} did not present all practical details of implementing the IBOSS procedure. For example, it is mentioned that if some data points have been included in the subdata by some covariate, the IBOSS algorithm requires to exclude them from consideration when using other covariates to select data points, but the implementation details were not provided. This step may add significant CPU time if not implemented appropriately. Thus, for completeness, we present a detailed IBOSS algorithm taking into account necessary practical considerations in Algorithm~\ref{alg:1}. 

\begin{algorithm}
  \caption{Original IBOSS algorithm}
  \label{alg:1}
  \begin{algorithmic}
    \STATE {\bf Input:} full data $\Fn$, subdata size $k$
    \STATE {\bf Output:} subdata set $\ds$
    \STATE $(N, p) = \textrm{size}(\cz)$
    \hfill \{get the full data size and covariate dimension\}
    \STATE $\S=\emptyset$
    \hfill \{initialize the index set of the subdata\}
    \STATE $r = \ceil{k/(2p)}$ 
    \hfill \{number of data points to take at each covariate tail\}
    \FOR {$j$ in $1, ..., p$}
    \STATE $\S$ = sort($\S$)
    \hfill \{sort elements in $\S$ in an increasing order\}
    \STATE $\u=(u_1, ..., u_{N-|\S|})\tp$
    \hfill \{allocate $\u$ to store elements of $\z_j$, $z_{ij},i\notin\S$\}
    \STATE $t = 1$, $s = 1$
    \hfill \{$t$ is the counter for $\u$; $s$ is the counter for $\S$\}
    \FOR {$i$ in $1, ..., N$}
    \IF {$i \neq \S[s]$}
    \STATE $u_t = z_{ij}$, $t=t+1$
    \hfill \{record $z_{ij}$ in $\u$ if $i\notin\S$\}
    \ELSIF {$s < |\S|$}
    \STATE $s=s+1$
    \hfill \{skip $i$ if it is already in $\S$\}
    \ENDIF
    \ENDFOR

    \STATE $l=u_{(r)}$, $q=u_{(N-r+1)}$
    \hfill \{using partition based quick selection\}
    \STATE $s=1$, $r_l=1$, $r_q=1$
    \hfill \{$r_l$, $r_q$ count numbers of data points in each tail\}
    \FOR {$i$ in $1, ..., N$}
    \IF {$|\S| \ge k$ or $(r_l > r$ and $r_q > r)$}
    \STATE break
    \hfill \{break when enough data points are taken\}
        \ELSIF {$s\le|\S|$ and $i=\S[s]$}
        \STATE         $s=s+1$
        \hfill \{skip $i$ if it is already in $\S$\}
        \ELSIF {$r_l\le r$ and $z_{ij}\le l$}
        \STATE $\S=\S\cup i$, $r_l=r_l+1$
        \hfill \{include one data point from the left tail\}
        \ELSIF {$r_q\le r$ and $z_{ij} \ge q$}
        \STATE $\S=\S\cup i$, $r_q=r_q+1$
        \hfill \{include one data point from the right tail\}
        \ENDIF
    \ENDFOR
    \ENDFOR

    \STATE $\S$=sort($\S$) 
    \STATE $\ds = \emptyset$
    \hfill \{initialize the set of the subdata\}
    \STATE $t=1$, $s=1$
    \FOR {$i$ in $1, ..., N$}
    \IF {$s\le |\S|$ and $i=\S[s]$}
    \STATE $\ds = \ds \cup (\z_i,y_i)$, $s=s+1$
    \hfill \{take the subdata according to $\S$\}
    \ENDIF
    \ENDFOR
    \RETURN $\ds$
  \end{algorithmic}
\end{algorithm}

\section{Divide-and-conquer IBOSS algorithm}
\label{sec:divide-conquer-iboss}

The IBOSS approach in Algorithm~\ref{alg:1} needs to look at the data column by column. Although for most programming languages, such as R \citep{R} and Julia \citep{julia}, data matrices are stored by columns in the memory, data files are often stored on hard drive by rows. Thus, if the data volume exceeds the size of the available memory, it is difficult to implement the IBOSS algorithm. To deal with the problem of limited memory, a natural approach is the divide-and-conquer procedure, in which the full data set is divided into partitions by rows and then each partition can be loaded into the memory and processed individually. To combine this procedure with the IBOSS procedure, we present the divide-and-conquer IBOSS method in Algorithm~\ref{alg:2}.

\begin{algorithm}
  \caption{Divide-and-conquer IBOSS algorithm}
  \label{alg:2}
  \begin{algorithmic}
    \STATE {\bf Input:} full data $\Fn$, subdata size $k$, partition size $n_B$.
    \STATE {\bf Output:} subdata set $\dsB$
    \IF {$r_B=k/(2pB)<1$}
    \PRINT ``The number of data points from each covariate tail is smaller than one.''
    \ENDIF
    \STATE ($\Fn^{(1)}$, ..., $\Fn^{(B)}$) = split($\Fn$, $n_B$)
    \hfill \{Divide the full data into $B$ partitions of\\ \hfill size $n_B$. This step can be skipped if the \\ \hfill full data are stored in multiple small files.\}
    \STATE $\dsB=\emptyset$
    \FOR {$b$ in $1, ..., B$}
    \STATE Run Algorithm~\ref{alg:1} on $\Fn^{(b)}$ with $k_b=\ceil{k/B}$ to obtain $\ds^{(b)}$
    \STATE $\dsB=\dsB\cup\ds^{(b)}$
    \ENDFOR
    \RETURN $\dsB$
  \end{algorithmic}
\end{algorithm}

\begin{remark}
  Algorithm~\ref{alg:2} takes a subdata $\ds^{(b)}$ from each partition $\Fn^{(b)}$ and then combines them to have a final subdata $\dsB$. This is recommended if only one machine is used, and the combined subdata should be small enough so that thorough analysis can be performed on the machine. If a distributed computing system is used and data partitions are processed in parallel by multiple machines, we can calculate the least squares estimators and their estimated variance-covariance matrices on each subdata $\ds^{(b)}$ in all machines and then aggregate these estimates using a linear combination with the inverses of the variance-covariance matrices as weights. The final estimators from these two approaches are identical \citep{LinXie2011,schifano2016online}. The latter procedure takes advantage of parallel computing facilities, but it has to carry out additional calculations on each machine if further regression diagnostics are to be performed. 
\end{remark}

\begin{remark}
  Using one machine to implement Algorithm~\ref{alg:2}, the step to divide the full data $\Fn$ into blocks $\Fn^{(1)}$, ..., $\Fn^{(B)}$ can be done by using the UNIX command \verb|split|. This command is also available for Windows through Cygwin. 
\end{remark}

\subsection{Theoretical properties}
In this section, we investigate the performance of Algorithm~\ref{alg:2} theoretically. With out loss of generality, we assume that $r_B=r/B=k/(2pB)$ and $n_B=N/B$ are both integers, where $B$ is the number of partitions on the full data.

Let $z_{(i)j}$ ($i=1, ..., N; j=1, ..., p$) be the $i$-th order statistic on the $j$-th covariate in the full data and $z_{b(i)j}$ ($b=1, ..., B; i=1, ..., n_B; j=1, ..., p$) be the $i$-th order statistics on the $j$-th covariate in the $b$-th block. 
Denote the covariate matrix for the subdata $\dsB$ by $\cz^*_{\dsB}$ and denote the sample correlation matrix of $\cz^*_{\ds}$ by $\mathbf{R}_{\dsB}$. Let $\cx^*_{\dsB}=(\mathbf{1},\cz^*_{\dsB})$.

For the divide-and-conquer IBOSS method, we first present a result showing the quality of using Algorithm~\ref{alg:2} to approximate $\zeta_N$, a typically unachievable upper bound of $|\M_{\bdelta}|$ defined in~\eqref{eq:15}.

\begin{theorem}\label{thm:3}
The subdata $\dsB$ selected using Algorithm~\ref{alg:2} satisfies that
\begin{equation}\label{eq:36}
  \frac{|(\cx^*_{\dsB})\tp\cx^*_{\dsB}|}{\zeta_N}
  \ge\max(C_R, C_E),
\end{equation}
where
\begin{align}
  C_R=&\frac{\lambda_{\min}^p(\bm{R}_{\dsB})}{(Bp)^p}
        \prod_{j=1}^p\left\{
        \sumb\left(\frac{z_{b(n_B-r_B+1)j}-z_{b(r_B)j}}
       {z_{(N)j}-z_{(1)j}}\right)^2\right\},\label{eq:16}\\
  C_E=&\frac{\lambda_{\min}^p(\bm{R}_{\dsB})}{(Bp)^p}
        \times\prod_{j=1}^p
        \left(\frac{z_{(N-r_B+1)j}-z_{(r_B)j}}{z_{(N)j}-z_{(1)j}}\right)^2,
        \label{eq:19}
\end{align}
and $\lambda_{\min}(\mathbf{R}_{\dsB})$ is the smallest eigenvalue of $\mathbf{R}_{\dsB}$.
\end{theorem}

\begin{remark}
  The lower bound of the approximation ratio in \eqref{eq:36} has two terms: $C_R$ in \eqref{eq:16} and $C_E$ in \eqref{eq:19}. Heuristically, $C_R$ is a bound corresponding to the scenario that the full data is divided randomly so that the covariate distributions for different partitions are the same. In this scenario $C_R$ is a sharper bound compared with $C_E$, i.e., $C_R>C_E$. On the other hand, $C_E$ is a bound corresponding to some extreme cases, e.g., ranges of covariate values for all blocks do not overlap. In this case, $C_E$ may be larger than $C_R$. 
\end{remark}

Theorem~\ref{thm:3} is aligned with Theorem 3 of \cite{WangYangStufken2018} for the original IBOSS algorithm, which shows that for the subdata $\ds$ obtained from Algorithm~\ref{alg:1}, the following inequality holds.
\begin{equation}\label{eq:7}
  \frac{|(\cx^*_{\ds})\tp\cx^*_{\ds}|}{\zeta_N}
  \ge\frac{\lambda_{\min}^p(\mathbf{R}_{\ds})}{p^p}
  \prod_{j=1}^p\left(\frac{z_{(N-r+1)j}-z_{(r)j}}{z_{(N)j}-z_{(1)j}}\right)^2,
\end{equation}
where $\cx^*_{\ds}=(\mathbf{1},\cz^*_{\ds})$, $\cz^*_{\ds}$ is the covariate matrix for the subdata $\ds$, and $\lambda_{\min}^p(\mathbf{R}_{\ds})$ is the minimum eigenvalue of $\mathbf{R}_{\ds}$, the correlation matrix of $\cz^*_{\ds}$. Comparing \eqref{eq:36} and \eqref{eq:7}, we see that there may be some information loss due to the divide-and-conquer procedure because $(z_{b(n_B-r_B+1)j}-z_{b(r_B)j})^2$ and $(z_{(N-r_B+1)j}-z_{(r_B)j})^2/B$ are typically smaller than $(z_{(N-r+1)j}-z_{(r)j})^2$. However, if $B$ is not too large,  $z_{b(n_B-r_B+1)j}-z_{b(r_B)j}$ and $z_{(N-r+1)j}-z_{(r)j}$ are on the same order under reasonable assumptions, so the lower bounds of the approximation ratios are at the same order. Theorem~\ref{thm:3} also indicates that $|(\cx^*_{\dsB})\tp\cx^*_{\dsB}|$ may achieve the same order of the unachievable upper bound $\zeta_N$ under some reasonable assumptions if $B$ and $p$ do not go to infinity.

Now we investigate the properties of the resulting estimator form Algorithm~\ref{alg:2}. 
Let $\hat\bbeta^{\dsB}=(\hat\beta_0^{\dsB}, \hat\beta_1^{\dsB}, ..., \hat\beta_p^{\dsB})\tp$ be the least squares estimator calculated from  the divide-and-conquer IBOSS subdata, namely,
\begin{equation*}
  \hat\bbeta^{\dsB}=\left(\sumb\sum_{i=1}^{n_B}
    \x_{bi}^*{\x_{bi}^*}\tp\right)^{-1}
  \left(\sumb\sum_{i=1}^{n_B}\x_{bi}^*y_{bi}^*\right),
\end{equation*}
where $\x_{bi}^*=(1,\z_{bi}^*)\tp$ and $(\z_{bi}^*,y_{bi}^*)$ is the $i$-th observation in the subdata $\dbs$ from the $b$-th block. 
To investigate the theoretical properties of $\hat\bbeta^{\dsB}$,
we focus on the variances because $\hat\bbeta^{\dsB}$ is unbiased for $\bbeta$.  

The following theorem provides finite sample bounds on variances of $\hat\beta_j^{\dsB}$, $j=0, 1, ..., p$. No quantity is required to go to infinity here. 

\begin{theorem}\label{thm:4}
If the sample correlation matrix for covariates $\cz^*_{\dsB}$ in the subdata $\dsB$ is column full rank, then the following results hold for the estimator, $\hat\bbeta^{\dsB}$, obtained from Algorithm~\ref{alg:2}:
\begin{align}
  &\Var(\hat\beta^{\dsB}_0|\cz)\ge\frac{\sigma^2}{k},
    \quad\text{ and}\label{eq:63}\\
 \frac{4\sigma^2}{k
  (z_{(N)j}-z_{(1)j})^2}
  \le&\Var(\hat\beta^{\dsB}_j|\cz)
  \le \min(V_{aj}, V_{ej}),\label{eq:64}
\end{align}
with probability one for $j=1, ..., p$, where
\begin{align*}
  V_{aj}&=\frac{4pB\sigma^2}
    {k\lambda_{\min}(\bm{R}_{\dsB})
    \sumb\left(z_{b(n_B-r_B+1)j}-z_{b(r_B)j}\right)^2},\\
  V_{ej}&=\frac{4pB\sigma^2}
    {k\lambda_{\min}(\bm{R}_{\dsB})(z_{(N-r_B+1)j}-z_{(r_B)j})^2},
\end{align*}
and $\lambda_{\min}(\bm{R}_{\dsB})$ is the minimum eigenvalue of the sample correlation matrix of $\cz^*_{\dsB}$. 
\end{theorem}

Theorem~\ref{thm:4} shows that for the intercept, the variance of the estimator $\hat\beta^{\dsB}_0$ is bounded from below by $\sigma^2/k$, which is the same as the bound for the original IBOSS algorithm. It indicates that this variance cannot go to zero for a fixed $k$. However, for slope parameters, the variances of the estimators $\hat\beta^{\dsB}_j$ may converge to zero under some assumptions even $k$ is fixed.

Theorem~\ref{thm:4} is aligned with Theorem 4 of \cite{WangYangStufken2018}, which shows that the variances of the slope estimators from the original IBOSS procedure are bounded from above by 
\begin{equation*}
  V_{oj}=\frac{4p\sigma^2}
   {k\lambda_{\min}(\mathbf{R}_{\ds})(z_{(N-r+1)j}-z_{(r)j})^2},
     \quad j=1, ..., p.
\end{equation*}
Comparing the upper bounds $V_{aj}$ and $V_{ej}$ with the upper bounds $V_{oj}$ for the original IBOSS procedure, we also see that the divide-and-conquer IBOSS algorithm may subject to some information loss. However, the information loss can be ignored asymptotically if the full data is partitioned randomly and the number of partitions does not go to infinity too fast. We derive the following theorem showing that the orders of $\Var(\hat\beta^{\dsB}_j|\cz)$ can be the same as those of the variances for the original IBOSS estimators. 

\begin{theorem}\label{thm:4-2}
Assume that covariate distributions are in the domain of attraction of the
generalized extreme value distribution, and  $\underset{N\rightarrow\infty}{\lim\inf}\lambda_{\min}(\bm{R}_{\dsB})>0$. For large enough $N$, when the full data is divided randomly in Algorithm~\ref{alg:2}, the following results hold for the estimator, $\hat\beta_j^{\dsB}$, $j=1, ..., p$.
\begin{align*}
 &\Var(\hat\beta^{\dsB}_j|\cz)=O_P\left\{
   \frac{p}{k(z_{(N)j}-z_{(1)j})^2}\right\},
   \quad j=1, ..., p,
\end{align*}
if one of the following conditions holds: 1)
$r$ and $B$ are fixed; 2) the support of $F_j$ is bounded 
and $r/N\rightarrow0$, where $F_j$ is the marginal distribution function of the $j$-th component of $\z$; 3) the upper endpoint for the support of $F_j$ is $\infty$ and the lower endpoint for the support of $F_j$ is finite, and 
\begin{equation}\label{eq:8}
  \frac{r}{N[1-F_j\{(1-\epsilon)F_j^{-1}(1-N^{-1})\}]}\rightarrow0
  \quad\text{ and }\quad
  \frac{F_j(1-1/N)}{F_j(1-B/N)}\rightarrow1,
\end{equation}
for all $\epsilon>0$; 4) the upper endpoint for the support of $F_j$ is finite and the lower endpoint for the support of $F_j$ is $-\infty$, and
\begin{equation}\label{eq:33}
  \frac{r}{NF_j\{(1-\epsilon)F_j^{-1}(N^{-1})\}}\rightarrow0,
  \quad\text{ and }\quad
  \frac{F_j(1/N)}{F_j(B/N)}\rightarrow1,
\end{equation}
for all $\epsilon>0$; and 5) the upper endpoint and the lower endpoint for the support of $F_j$ are $\infty$ and $-\infty$, respectively, and both \eqref{eq:8} and \eqref{eq:33} hold. 
\end{theorem}

The convergence rates of $\Var(\hat\beta^{\dsB}_j|\cz)$ described in Theorem~\ref{thm:4-2} are the same as those given in Theorem 5 of \cite{WangYangStufken2018} for the original IBOSS estimators of slope parameters.

For case 1), if $p$ is fixed, then the total subdata size $k$ is also fixed. As long as the supports of covariate distributions are not bounded, $z_{(N-r+1)j}-z_{(r)j}\rightarrow\infty$ in probability as $n\rightarrow\infty$, and thus the variances $\Var(\hat\beta^{\dsB}_j|\cz)$ converge to zero in probability. This is not the case for random subsampling-based methods \citep{PingMa2015-JMLR, WangZhuMa2017}. In this case, the asymptotic expression for the variance-covariance matrix of $\hat\bbeta^{\dsB}$ when $\z_i$ follow normal and lognormal distributions are identical to those in Theorem 6 of \cite{WangYangStufken2018}. 
For case 2), $z_{(N-r+1)j}-z_{(r)j}$ goes to some finite constant, so it is necessary that $k\rightarrow\infty$ in order for $\Var(\hat\beta^{\dsB}_j|\cz)$ to converge to zero. For case 3), the condition in \eqref{eq:8} impose constrains on $r$, $B$, $N$, and the tail behavior of the covariate distributions. For example, if
\begin{equation}\label{eq:20}
  F_j(z)=1-\exp\{-z^\gamma h(z)\},
\end{equation}
where $\gamma>0$ and $h(z)$ is a slowly varying function, then the first constrain in \eqref{eq:8} is to require that $\log r/\log N\rightarrow0$ \citep{hall1979relative}. The distribution in~\eqref{eq:20} includes many commonly seen distributions, such as exponential distribution, gamma distribution, Gumbel distribution, Laplace distribution, and normal distribution. For these distributions, if $\log B/\log N\rightarrow0$, then the second constrain in \eqref{eq:8} also holds. For case 4), the condition in \eqref{eq:33} is essentially the same as that in~\eqref{eq:8} if one takes transformation $\z=-\z$. For case 5, it is a combination of cases 3 and 4. Under the given conditions, $F_j^{-1}(1-N^{-1})/z_{(N)j}\rightarrow1$ and $F_j^{-1}(N^{-1})/z_{(1)j}\rightarrow1$, so the rates for 
$\Var(\hat\beta^{\dsB}_j|\cz)$ to converge to zero are 
\begin{align*}
  \frac{p}{k\big\{F_j^{-1}(1-N^{-1})-F_j^{-1}(N^{-1})\big\}^2}, \quad
  j=1, ..., p.
\end{align*}

\section{Numerical experiments}
\label{sec:numer-exper}
In this section, we use numerical experiments to evaluate the empirical performance of the divide-and-conquer IBOSS algorithm.

We generate data from model~\eqref{eq:14} with $\beta_0=1$ and $\bbeta_1$ being a 50 dimensional vector of ones. We assume that $\varepsilon_i$ are independent following $N(0,1)$. The following five cases are considered to generate the covariate matrix $\cz$.

\begin{enumerate}[{Case }1:, itemindent=0.7cm, leftmargin=*]
\item \textbf{Normal}. We generate $\cz$ from a multivariate normal distribution, $N(\bm{0}, \bSigma)$, where the $(i,j)$-th element of $\bSigma$ is 0.5 if $i\neq j$ and 1 otherwise. 
\item \textbf{LogNormal}. We generate $\cz$ from a multivariate lognormal distribution, namely, generate $\mathcal{V}$ from $N(\bm{0}, \bSigma)$ as defined in Case 1 and then set $\cz=e^{\mathcal{V}}$, where the exponentiation is element-wise.
\item $\bm{T_2}$. We generate $\cz$ from a multivariate $t$ distribution with two degrees of freedom $t_2(\bm{0}, \bSigma)$ with $\bSigma$ being the same as in Case 1.
\item \textbf{Mix with order}. We generate $\cz$ from five different distributions. Specifically, generate $\cz_1$ from $N(\bm{0}, \bSigma)$; generate $\cz_2$ from $t_2(\bm{0}, \bSigma)$; generate $\cz_3$ from $t_3(\bm{0}, \bSigma)$; generate $\cz_4$ with its elements independently following the same uniform distribution between zero and two; and generate $\cz_5$ from the multivariate lognormal distribution defined in Case 2. Here, $\cz_i$ ($i=1, ..., 5$) all contain $n/5$ rows, and $\bSigma$ is the same as defined in Case 1. Set $\cz=(\cz_1\tp,\cz_2\tp,\cz_3\tp,\cz_4\tp,\cz_5\tp)\tp$. 
\item \textbf{Mix random order}. We generate $\cz$ using the same procedure as in Case 4, and then we randomize the row orderings. 
\end{enumerate}

Here, Cases 1-4 are the same as those in \cite{WangYangStufken2018}. Cases 4 and 5 would produce identical results for the full data analysis approach but produce different results for the divide-and-conquer IBOSS algorithm as different row orderings will affect the data partitions. The purpose of considering both Cases 4 and 5 is to investigate the effect of different data partitions on the the divide-and-conquer IBOSS algorithm. We divide the full data according to the sequence of row numbers in the numerical experiments.

We repeat the simulation for $T=1000$ times to calculate the empirical mean squared error (MSE) as MSE=$T^{-1}\sum_{t=1}^T\|\bbeta_1^{(t)}-\bbeta_1\|^2$ for the slope parameter, where $\bbeta_1^{(t)}$ is the estimate at the $t$-th repetition.

\begin{figure}
  \centering
  \begin{subfigure}{0.49\textwidth}
    \includegraphics[width=\textwidth,page=1]{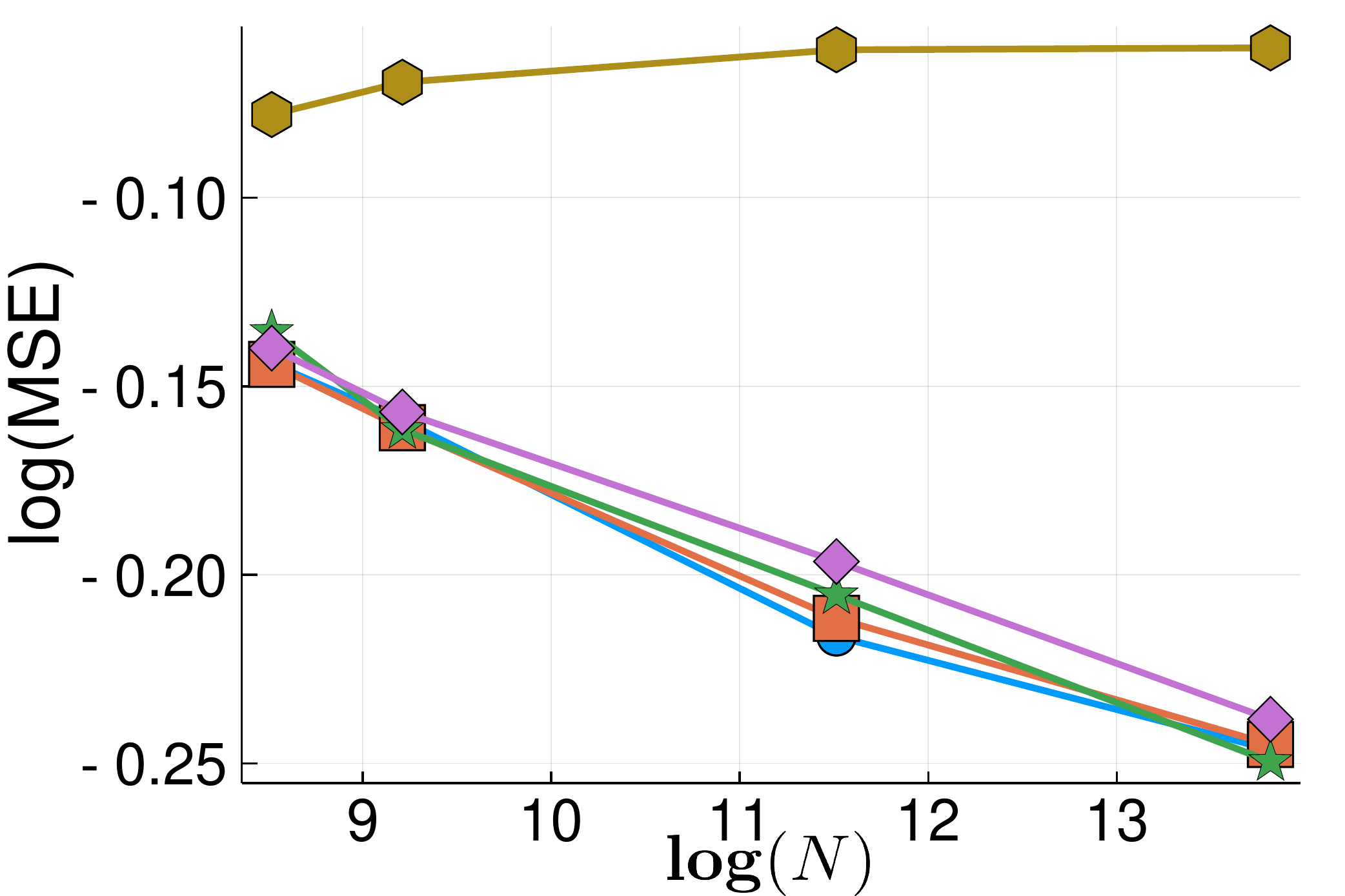}
    \caption{Case 1: \textbf{Normal}}
  \end{subfigure}
  \begin{subfigure}{0.49\textwidth}
    \includegraphics[width=\textwidth,page=2]{mse.pdf}
    \caption{Case 2: \textbf{LogNormal}}
  \end{subfigure}
  \begin{subfigure}{0.49\textwidth}
    \includegraphics[width=\textwidth,page=3]{mse.pdf}
    \caption{Case 3: $\bm{T_2}$}
  \end{subfigure}
  \begin{subfigure}{0.49\textwidth}
    \includegraphics[width=\textwidth,page=4]{mse.pdf}
    \caption{Case 4: \textbf{Mix with order}}
  \end{subfigure}
  \begin{subfigure}{0.49\textwidth}
    \includegraphics[width=\textwidth,page=5]{mse.pdf}
    \caption{Case 5: \textbf{Mix random order}}
  \end{subfigure}
  \begin{subfigure}{0.49\textwidth}
    \includegraphics[width=\textwidth,page=6]{mse.pdf}
  \end{subfigure}
  \caption{MSE of the slope parameter against full data sample size for the five cases of covariate distributions. The subdata
    size $k$ is fixed at $k=1000$.}
  \label{fig:1}
\end{figure}

Figure~\ref{fig:1} plots log of empirical MSEs against full data sample size $N$ of $N= 5\times10^3, 10^4, 10^5$ and $10^6$, with total subdata size $k=1000$. Here the natural logarithm is taken to have better presentations. For the number of partitions $B$, we considered four values: $B=1, 2, 5$, and $10$, and the corresponding values of $r_B$ are $r_B=10, 5, 2$, and $1$, respectively. Note that when $B=1$, the method is the original IBOSS approach. For comparison, the uniform Poisson subsampling method is also implemented. From Figure~\ref{fig:1}, the empirical MSE decreases as $N$ gets large for the divide-and-conquer IBOSS approach, while it stays constant for the random sampling approach. Different values of $B$ have some effect on the performance of divide-and-conquer IBOSS approach, the effect is not very significant, especially for Cases 2, 3, and 5. Comparing results in Cases 4 and 5, we see that different row orderings affect the performance of the divide-and-conquer IBOSS algorithm, and the random row ordering (random partitions) is preferable. 

\begin{figure}
  \centering
  \begin{subfigure}{0.49\textwidth}
    \includegraphics[width=\textwidth,page=1]{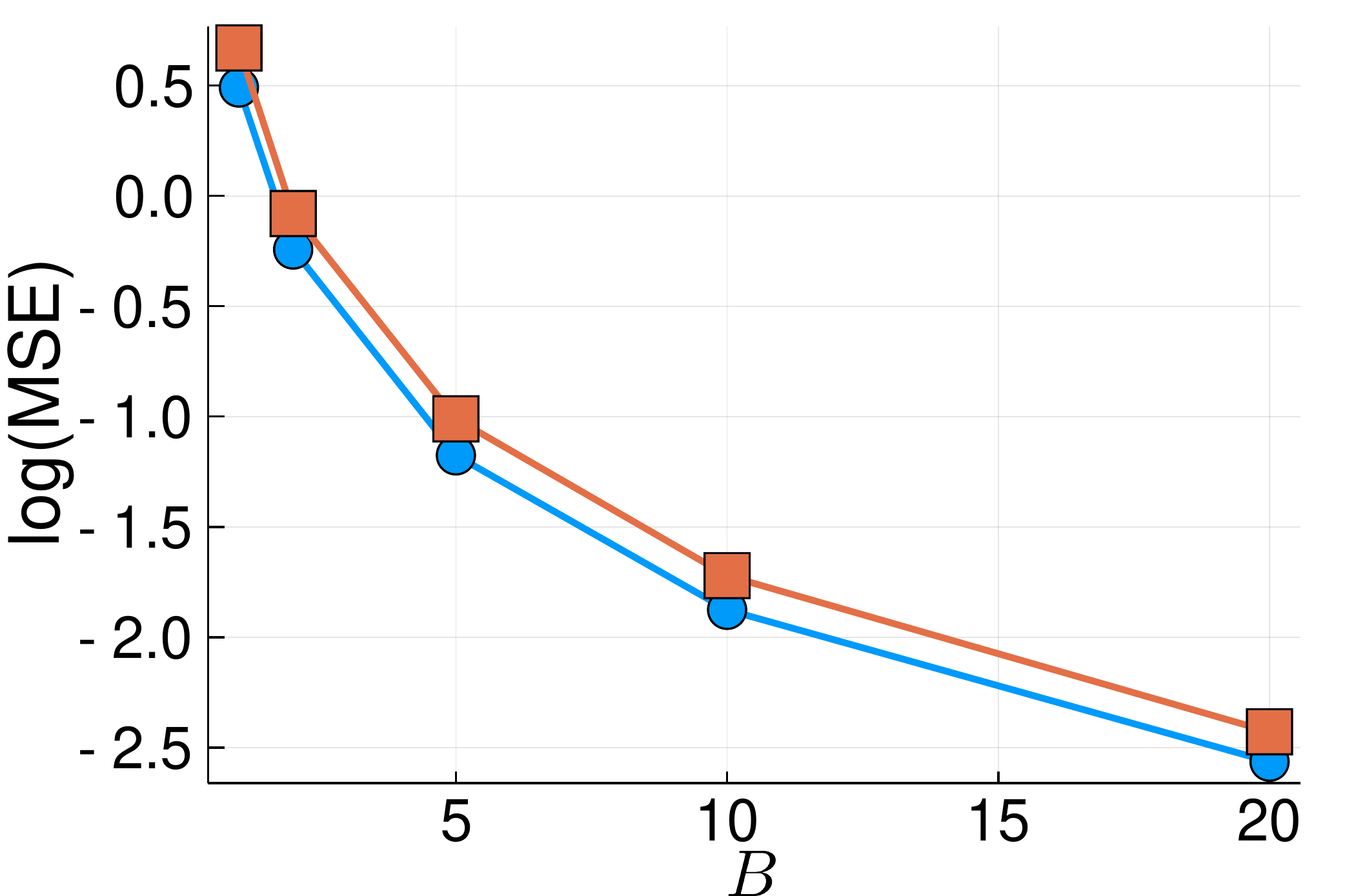}
    \caption{Case 1: \textbf{Normal}}
  \end{subfigure}
  \begin{subfigure}{0.49\textwidth}
    \includegraphics[width=\textwidth,page=2]{mse_Bc.pdf}
    \caption{Case 2: \textbf{LogNormal}}
  \end{subfigure}
  \begin{subfigure}{0.49\textwidth}
    \includegraphics[width=\textwidth,page=3]{mse_Bc.pdf}
    \caption{Case 3: $\bm{T_2}$}
  \end{subfigure}
  \begin{subfigure}{0.49\textwidth}
    \includegraphics[width=\textwidth,page=4]{mse_Bc.pdf}
    \caption{Case 4: \textbf{Mix with order}}
  \end{subfigure}
  \begin{subfigure}{0.49\textwidth}
    \includegraphics[width=\textwidth,page=5]{mse_Bc.pdf}
    \caption{Case 5: \textbf{Mix random order}}
  \end{subfigure}
  \begin{subfigure}{0.49\textwidth}
    \includegraphics[width=\textwidth,page=6]{mse_Bc.pdf}
  \end{subfigure}
  \caption{MSE of the slope parameter against number of partitions $B$ with a fixed $r_B$. The subdata size $k$ changes as $B$ changes.}
  \label{fig:2}
\end{figure}

We also consider the scenario that $r_B$ is fixed while $B$ changes. This is to mimic the scenario that one may have $B$ machines to use while the computational capacity for each machine is limited. This is often the case for parallel distributed computing systems. We set $r_B=5$ and choose $B$ to be $B=1$, 2, 5, 10, and 20, which gives the values of $k$ as $k=500$, 1000, 2500, 5000, and 10000, respectively. We fixed $N=10^6$.  Figure~\ref{fig:2} presents the results. It is seen that both the divide-and-conquer method and the Poisson subsampling method improve as $B$ increases because the total subdat size $k$ increases. The divide-and-conquer method dominates the Poisson subsampling method and the difference in their performance depends on the covariate distribution. 

\begin{figure}
  \centering
  \begin{subfigure}{0.49\textwidth}
    \includegraphics[width=\textwidth,page=1]{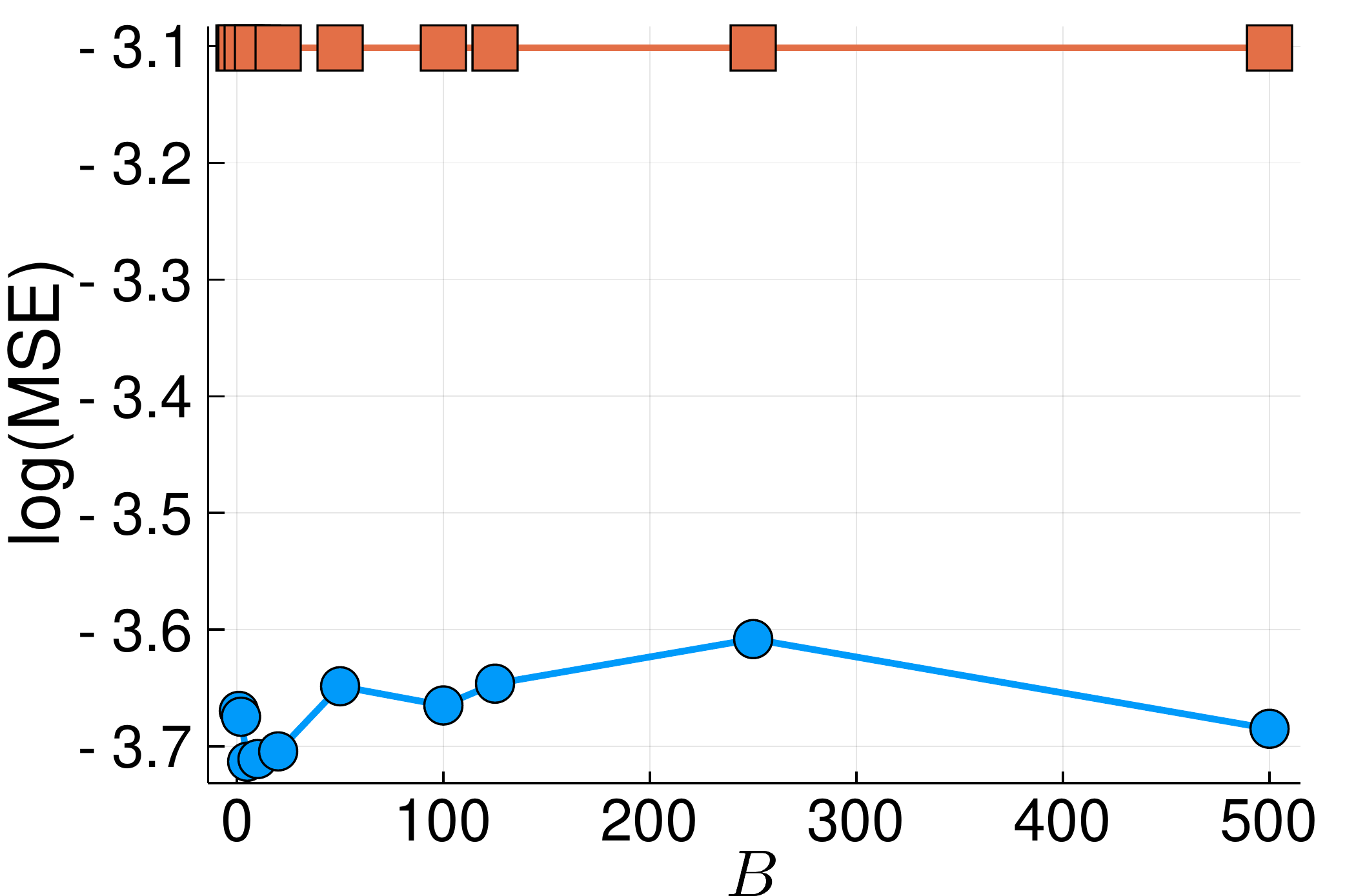}
    \caption{Case 1: \textbf{Normal}}
  \end{subfigure}
  \begin{subfigure}{0.49\textwidth}
    \includegraphics[width=\textwidth,page=2]{mse_kfBc.pdf}
    \caption{Case 2: \textbf{LogNormal}}
  \end{subfigure}
  \begin{subfigure}{0.49\textwidth}
    \includegraphics[width=\textwidth,page=3]{mse_kfBc.pdf}
    \caption{Case 3: $\bm{T_2}$}
  \end{subfigure}
  \begin{subfigure}{0.49\textwidth}
    \includegraphics[width=\textwidth,page=4]{mse_kfBc.pdf}
    \caption{Case 4: \textbf{Mix with order}}
  \end{subfigure}
  \begin{subfigure}{0.49\textwidth}
    \includegraphics[width=\textwidth,page=5]{mse_kfBc.pdf}
    \caption{Case 5: \textbf{Mix random order}}
  \end{subfigure}
  \begin{subfigure}{0.49\textwidth}
    \includegraphics[width=\textwidth,page=6]{mse_kfBc.pdf}
  \end{subfigure}
  \caption{MSE of the slope parameter against numbers of partitions $B$ with a fixed $k=2000$ for the five cases of covariate distributions. Here, $r_B$ changes as $B$ changes.}
  \label{fig:3}
\end{figure}

To further explore the effect of larger number of partitions $B$ with a fixed $k=2000$, we reduce the value of $p$ to be $p=2$ so that the maximum value of $B$ is $B=500$ corresponding to $r_B=1$. We consider $B=1$, 2, 5, 10, 20, 50, 100, 125, 250, and 500, which gives values of $r_B$ as $r_B=500$, 250, 100, 50, 25, 10, 5, 4, 2, and 1, respectively. Figure~\ref{fig:3} gives the results. We see that the effect of $B$ on the performance of the divide-and-conquer IBOSS algorithm is small compared with its significant improvement relative to the Poisson subsampling method. 
Additionally, there is no clear increasing pattern for the empirical MSE as $B$ gets large. This is different from the observations for regular cases in the existing literature where the estimation efficiency of the divide-and-conquer estimators often decreases as the number of partitions increases \citep[e.g.][]{LinXie2011, schifano2016online, battey2018distributed}. 

We investigate the computing times of the divide-and-conquer IBOSS algorithm for the case that data blocks are stored on hard drive. We set $N=10^7$, $p=100$, $k=10^4$, and choose values of $B$ to be $B=2, 5, 10, 25$, and $50$, giving values of $r_B=25, 10,  5,  2$, and $1$, respectively. 
For comparisons, we implement the full data divide-and-conquer approach in which all observations in each block of data are used to calculate least squares estimators and these estimators are then aggregated to form a full data estimator. We also implement the Poisson subsampling method on each data block to obtain subdata sets and then combine these subdata sets to calculate an estimate. Table~\ref{tab:1} presents the results on computing times, where the time to load each data block into the memory is also counted. The calculation was performed on a computer with Intel Core i7 processors, a 64 GB RAM, and SSD hard drives, running Ubuntu 18.04 Linux system. Only one CPU was used for fair comparisons. 
It is seen that the divide-and-conquer IBOSS algorithm is faster than the full data approach, but is not as fast as the the uniform subsampling approach. For the divide-and-conquer IBOSS approach and the full data approach, as the number of blocks increases, the computing times are reduced. However, this is not the case for the uniform subsampling approach, for which the computing time decreases first but then increases as $B$ increases. This is because as $B$ gets large, the data loading time becomes the dominating term for the time required by the divide-and-conquer uniform Poisson subsampling method. 

\begin{table}
  \caption{Computing times for different number of partitions $B$ with $N=10^7$, $p=10^2$, and $k=10^4$. ``FULL'' uses all the observations in each data block; ``IBOSS'' uses the IBOSS algorithm to select a subdata set of $k/B$ observations; and ``UNI uses Poisson subsampling to select a subdata set of $k/B$ observations.}
  \label{tab:1}
  \centering
\begin{tabular}{|c|ccccc|}
\hline
  Methods  & $B=2$ & $B=5$ & $B=10$ & $B=25$ & $B=50$ \\
  \hline
  FULL  & 98.80 & 95.76 & 90.82 & 86.41 & 85.28 \\
   IBOSS & 29.74 & 24.90 & 20.79 & 18.81 & 18.42 \\
  UNI & 6.81 & 5.67 & 4.32 & 6.75 & 14.08 \\
  \hline
\end{tabular}
\end{table}

\section{Concluding remarks}
\label{sec:concluding-remarks}
In this paper, we have developed a divide-and-conquer IBOSS method in the context of linear regression to solve the issue that the data volume is too large to apply the D-optimality motivated IBOSS method in memory. We have derived theoretical guarantees for the resultant estimator and drawn comparisons with the original IBOSS approaches. We have also carried out numerical experiments to evaluate and demonstrate the proposed method.

We found that the estimation efficiency of the divide-and-conquer IBOSS algorithm may not be a monotonic function of the number of partitions as in other divide-and-conquer methods, which indicates the possibility of a nontrivial optimal number of partitions. This is an interesting and challenging question warrants for further investigations. 

\section{Proofs}
\subsection{Proof of Theorem~\ref{thm:3}}
\begin{proof} 
  For $l\neq j$, let $z_j^{(i)l}$ be the concomitant of $z_{(i)l}$ for $z_j$, i.e., if $z_{(i)l}=z_{sl}$ then $z_j^{(i)l}=z_{sj}$, $i=1, ..., N$. 
  Let $\bar{z}_{j\dsB}^*$ and $\vr(z_{j\dsB}^*)$ be the sample mean and sample variance for covariate $z_j$ of subdata $\dsB$.
  For the proof of Theorem 3 in \cite{WangYangStufken2018}, we know that
\begin{align}\label{eq:34}
  |(\cx^*_{\dsB})\tp\cx^*_{\dsB}|
  \ge k(k-1)^p\lambda_{\min}^p(\bm{R}_{\dsB})\prod_{j=1}^p\vr(z_{j\dsB}^*).
\end{align}
Let $k_B=k/B$ be the number of data points to take from each partition, and $z_{bij}^*$ be the $i$-th observation on the $j$-th covariate in the subdata $\dbs$ from the $b$-th partition. 
The sample variance for each $j$ satisfies,  
\begin{align*}
  &(k-1)\vr(z_{j\dsB}^*)
  =\sumk\left(z_{ij}^*-\bar{z}_{j\dsB}^*\right)^2
  =\sumb\sumkb\left(z_{bij}^*-\bar{z}_{j\dsB}^*\right)^2\notag\\
  \ge&\sumb\left(\sumrb+\sum_{i=n_B-r_B+1}^{n_B}\right)
       \left(z_{b(i)j}-\bar{z}_{bj}^{**}\right)^2\notag\\
  =&\sumb\Bigg\{\sumrb\left(z_{b(i)j}-\bar{z}_{bj}^{*l}\right)^2
     +\sum_{i=n_B-r_B+1}^{n_B}\left(z_{b(i)j}-\bar{z}_{bj}^{*u}\right)^2
     +\frac{r_B}{2}\left(\bar{z}_{bj}^{*u}-\bar{z}_{bj}^{*l}\right)^2
     \Bigg\}\notag\\
  \ge&\frac{r_B}{2}
       \sumb\left(\bar{z}_j^{*u}-\bar{z}_j^{*l}\right)^2
  \ge\frac{r_B}{2}\sumb\left(z_{b(n_B-r_B+1)j}-z_{b(r_B)j}\right)^2,
\end{align*}
where $\bar{z}_{bj}^{**}=\left(\sumrb+\sum_{i=n_B-r_B+1}^{n_B}\right)z_{b(i)j}/(2r_B)$, $\bar{z}_{bj}^{*l}=\sumrb z_{b(i)j}/{r_B}$,\\ and $\bar{z}_{bj}^{*u}=\sum_{i=n_B-r_B+1}^{n_B} z_{b(i)j}/(r_B)$.
Thus,
\begin{align}\label{eq:12}
  \vr(z_{j\dsB}^*)
  \ge&\frac{r(z_{(N)j}-z_{(1)j})^2}{2B(k-1)}
       \sumb\left(\frac{z_{b(n_B-r_B+1)j}-z_{b(r_B)j}}
       {z_{(N)j}-z_{(1)j}}\right)^2,
\end{align}
which, combined with \eqref{eq:34}, shows that
\begin{align*}
  |(\cx^*_{\dsB})\tp\cx^*_{\dsB}|
  \ge& \frac{r^p}{2^pB^p}
     k\lambda_{\min}^p(\bm{R}_{\dsB})\prod_{j=1}^p(z_{(N)j}-z_{(1)j})^2\\
  &\qquad   \times\prod_{j=1}^p
       \sumb\left(\frac{z_{b(n_B-r_B+1)j}-z_{b(r_B)j}}
       {z_{(N)j}-z_{(1)j}}\right)^2.
\end{align*}
This shows that
\begin{align}\label{eq:10}
  \frac{|(\cx^*_{\dsB})\tp\cx^*_{\dsB}|}{M_N}
  \ge&\frac{\lambda_{\min}^p(\bm{R}_{\dsB})}{(Bp)^p}
       \prod_{j=1}^p\sumb\left(\frac{z_{b(n_B-r_B+1)j}-z_{b(r_B)j}}
       {z_{(N)j}-z_{(1)j}}\right)^2.
\end{align}

Each numerator in the summation of the bound in \eqref{eq:10} relay on the covariate range of each data partition. 
If the full data is not divided randomly, this may not produce a sharp bound and using the full data covariate ranges may produce a better bound. We use this idea to derive the bound $C_E$ in the following. From Algorithm~\ref{alg:2}, for each $j=1, ..., p$, the $r_B$ data points with the smallest value of $z_j$ and the $r_B$ data points with the largest value of $z_j$ must be included in $\dsB$. Thus, for each sample variance, 
\begin{align*}
  (k-1)\vr(z_{j\dsB}^*)
  \ge&\left(\sumrb+\sumrbN\right)
       \left(z_{(i)j}-\bar{z}_{j\dsB}^*\right)^2\notag\\
  \ge&\left(\sumrb+\sumrbN\right)
       \left(z_{(i)j}-\bar{z}_j^{**}\right)^2\notag\\
  \ge&\frac{r_B}{2}\left(\bar{z}_j^{*u}-\bar{z}_j^{*l}\right)^2
  \ge\frac{r_B}{2}\left(z_{(N-r_B+1)j}-z_{(r_B)j}\right)^2.
\end{align*}
In this case,  $\bar{z}_j^{**}=(\sumrb+\sumrbN)z_{(i)j}/(2r_B)$, $\bar{z}_j^{*l}=\sumr z_{(i)j}/(r_B)$,\\ and $\bar{z}_j^{*u}=\sumrn z_{(i)j}/(r_B)$. 
Thus,
\begin{align}
  \vr(z_{j\dsB}^*)\ge
  &\frac{r(z_{(N)j}-z_{(1)j})^2}{2(k-1)B}
    \left(\frac{z_{(N-r_B+1)j}-z_{(r_B)j}}
    {z_{(N)j}-z_{(1)j}}\right)^2. \label{eq:71}
\end{align}
This, combined with \eqref{eq:34}, shows that
\begin{align}\label{eq:11}
  \frac{|(\cx^*_{\dsB})\tp\cx^*_{\dsB}|}{M_N}
  \ge&\frac{\lambda_{\min}^p(\bm{R}_{\dsB})}{(Bp)^p}
  \times\prod_{j=1}^p
     \left(\frac{z_{(N-r_B+1)j}-z_{(r_B)j}}{z_{(N)j}-z_{(1)j}}\right)^2.
\end{align}
The proof finishes if we combine \eqref{eq:10} and \eqref{eq:11}. 
\end{proof}

\subsection{Proof of Theorem~\ref{thm:4}}
\begin{proof}
  The proof for \eqref{eq:63} is similar to the proof of inequality (19) in \cite{WangYangStufken2018}. 
 For \eqref{eq:64}, from the proof of Theorem 4 in \cite{WangYangStufken2018},
we know that
\begin{align}\label{eq:3}
  \vr(z_{j\dsB}^*)&\le\frac{k}{4(k-1)}\left(z_{(N)j}-z_{(1)j}\right)^2,
\end{align}
and
\begin{equation}\label{eq:62}
  \Var(\hat\beta^{\dsB}_j|\cz)
  =\frac{\sigma^2}{k-1}\frac{(\bm{R}_{\dsB}^{-1})_{jj}}{\vr(z_{j}^*)}.
\end{equation}
From the \eqref{eq:3}, \eqref{eq:62}, and the fact that
$(\bm{R}_{\dsB}^{-1})_{jj}\ge1$, we have
\begin{equation}\label{eq:6}
  \Var(\hat\beta^{\dsB}_j|\cz)
  \ge\frac{4\sigma^2(z_{(N)j}-z_{(1)j})^2}{k}.
\end{equation}
From \eqref{eq:12}, \eqref{eq:71} and the fact that $(\bm{R}_{\dsB}^{-1})_{jj}\le\lambda_{\min}^{-1}(\bm{R}_{\dsB})$, we have
\begin{align}\label{eq:4}
  \Var(\hat\beta^{\dsB}_j|\cz)
  &\le\frac{4pB\sigma^2}
    {k\lambda_{\min}(\bm{R}_{\dsB})(z_{(N-r_B+1)j}-z_{(r_B)j})^2},
  \quad\text{ and}\\
  \Var(\hat\beta^{\dsB}_j|\cz)
  &\le\frac{4pB\sigma^2}
    {k\lambda_{\min}(\bm{R}_{\dsB})
    \sumb\left(z_{b(n_B-r_B+1)j}-z_{b(r_B)j}\right)^2}.\label{eq:13}
\end{align}
The proof finishes by combining \eqref{eq:6}, \eqref{eq:4}, and \eqref{eq:13}.
\end{proof}

\subsection{Proof of Theorem~\ref{thm:4-2}}
\begin{proof}
  For the first case that $B$ and $r$ are fixed, from \eqref{eq:64} with $V_{ej}$, we only need to verify that
  \begin{equation}\label{eq:69}
    \frac{z_{(N)j}-z_{(1)j}}{z_{(N-r_B+1)j}-z_{(r_B)j}}=\Op,
  \end{equation}
  which is true according to Theorems 2.8.1 and 2.8.2 in \cite{galambos1987asymptotic}.

  For the second case, $z_{b(n_B-r_B+1)j}-z_{b(r_B)j}$ and $z_{(N)j}-z_{(1)j}$ converge to the same fixed constant in probability, and $z_{b(n_B-r_B+1)j}-z_{b(r_B)j}$ are bounded by this constant for all $b$. Thus,
\begin{align*}
    \frac{1}{B}\sumb(z_{b(n_B-r_B+1)j}-z_{b(r_B)j})^2
\end{align*}
converges to the same finite constant. Thus, \eqref{eq:69} can be easily verified.

For the third case, let $g_{N,j}=F_j^{-1}(1-1/N)$ and $g_{n_B,j}=F_j^{-1}(1-1/n_B)$. When \eqref{eq:8} holds, from the proof of Theorem 5 in \cite{WangYangStufken2018}, we have $z_{(N)j}/g_{N,j}=1+\op$ and $z_{(n_B-r_B+1)j}/g_{n_B,j}=1+\op$. Thus, noting that $z_{b(r)j}$ and $z_{(1)j}$ are bounded in probability when the lower endpoint for the support of $F_j$ is finite, we have
\begin{align*}
  \frac{z_{b(n_B-r_B+1)j}-z_{b(r)j}}{z_{(N)j}-z_{(1)j}}=1+\op.
\end{align*}
Note that  $\left|\frac{z_{b(n_B-r_B+1)j}-z_{b(r)j}}{z_{(N)j}-z_{(1)j}}\right|$ are bounded by the same constant for all $b$, so 
\begin{align*}
  \frac{1}{B}\sumb\frac{z_{b(n_B-r_B+1)j}-z_{b(r)j}}{z_{(N)j}-z_{(1)j}}=1+\op.
\end{align*}
Combining this and \eqref{eq:64} with $V_{aj}$, the result follows. 

For the forth case, the result follows from reversing the signs of the covariates in the proof for the third case. 

The proof for the fifth case is obtained by combining the proof for the third case and the proof for the fourth case.
\end{proof}

\singlespacing


\end{document}